\documentclass[lettersize,journal]{IEEEtran}
\usepackage{amsmath,amsfonts}
\usepackage{algorithmic}
\usepackage{algorithm}
\usepackage{array}
\usepackage[caption=false,font=normalsize,labelfont=sf,textfont=sf]{subfig}
\usepackage{textcomp}
\usepackage{stfloats}
\usepackage{url}
\usepackage{verbatim}

\usepackage{multirow}
\usepackage{graphicx}
\usepackage{cite}
\usepackage{color}
\usepackage{soul}
\begin{document}

\title{\huge Synergizing AI and Digital Twins for Next-Generation Network Optimization, Forecasting, and Security}

\author{Zifan Zhang, Minghong Fang,~\IEEEmembership{Member,~IEEE,} Dianwei Chen, \\
Xianfeng Yang,~\IEEEmembership{Member,~IEEE,} Yuchen Liu,~\IEEEmembership{Member,~IEEE}




\thanks{
Z. Zhang and Y. Liu are with the Department of Computer Science, North Carolina State University, Raleigh, NC, 27695, USA (Email: \{zzhang66, yuchen.liu\}@ncsu.edu).
\textit{(Corresponding author: Yuchen Liu.)}}
\thanks{M. Fang is with the Department of Computer Science and Engineering, University of Louisville, Louisville, KY, 40208, USA (Email: \{ minghong.fang@louisville.edu)\}.}
\thanks{
D. Chen and X. Yang are with the Department of Civil and Environmental Engineering at the University of Maryland, College Park, MD, 20742, USA (Email: \{dwchen98, xtyang\}@umd.edu).
}

}

\maketitle

\begin{abstract}

Digital network twins (DNTs) are virtual representations of physical networks, designed to enable real-time monitoring, simulation, and optimization of network performance. 
When integrated with machine learning (ML) techniques, particularly federated learning (FL) and reinforcement learning (RL), DNTs emerge as powerful solutions for managing the complexities of network operations.
This article presents a comprehensive analysis of the synergy of DNTs, FL, and RL techniques, showcasing their collective potential to address critical challenges in 6G networks.
We highlight key technical challenges that need to be addressed, such as ensuring network reliability, achieving joint data-scenario forecasting, and maintaining security in high-risk environments. Additionally, we propose several pipelines that integrate DNT and ML within coherent frameworks to enhance network optimization and security. 
Case studies demonstrate the practical applications of our proposed pipelines in edge caching and vehicular networks. In edge caching, the pipeline achieves over 80\% cache hit rates while balancing base station loads. In autonomous vehicular systems, it ensures a 100\% no-collision rate, showcasing its reliability in safety-critical scenarios.
By exploring these synergies, we offer insights into the future of intelligent and adaptive network systems that automate decision-making and problem-solving.

\end{abstract}

\begin{IEEEkeywords}
Digital twins, artificial intelligence, federated learning, reinforcement learning, security
\end{IEEEkeywords}


\section{Introduction}

Recently, digital network twins (DNTs) have been transforming the management and optimization of networks of 6G and beyond. Unlike traditional digital twins, which typically focus on single systems, DNTs excel in coordinating interactions between multiple twins within a network, ensuring efficient data flow, resilience, and adaptability across the entire communication infrastructure~\cite{tang2022survey}. As an integrated simulator, predictor, and optimizer, DNT serves as a real-time, interactive virtual platform for monitoring, simulating, and enhancing network performance. These virtual models can represent both the frontend infrastructure and underlying operational process, including device properties, system inter-relationships, and user behavior, enabling operators to predict potential issues and test new configurations prior to deployment.

\begin{figure*}[htbp]
    \centering
    \includegraphics[width=0.95\textwidth]{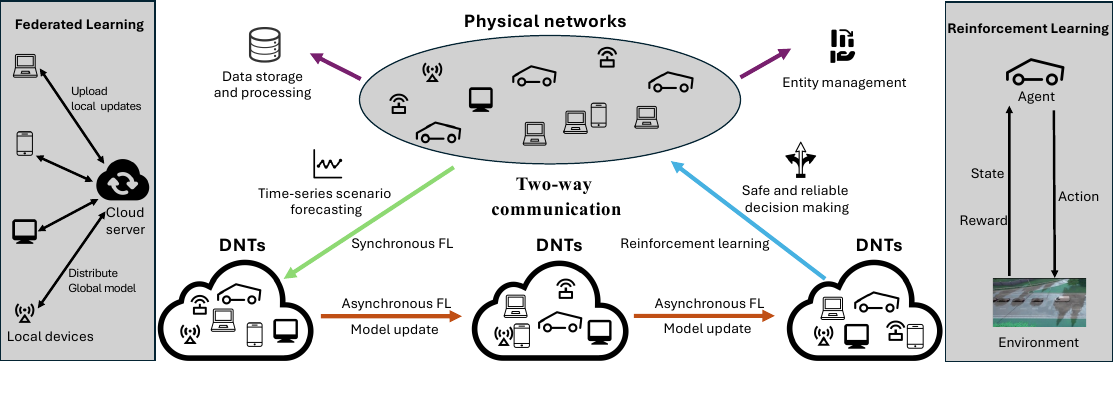}
    \vspace{-0.1in}
    \caption{Framework of the collaboration of DNT and AI.}
    \label{fig:framework}
\end{figure*}

As the core enabler, machine learning (ML), particularly federated learning (FL) and reinforcement learning (RL), significantly enhances the functionality and effectiveness of DNTs. 
Specifically, FL involves collaborative \textit{interactions among multiple agents} in parallel, where each contributing local model updates, while RL focuses on the decision-making process of a single agent \textit{interacting with its environment}. As shown in Fig.~\ref{fig:framework}, FL enables distributed model training by allowing devices to process data locally and share only model updates, safeguarding data privacy, reducing communication overhead, and enabling continuous model evolution~\cite{wu2021digital}. This empowers DNTs to rapidly adapt to changing network conditions without risking privacy or overwhelming infrastructure resources. RL, on the other hand, excels in dynamic, uncertain environments by learning optimal policies from interactions with the network, emphasizing its role as a model for agent-environment interaction, allowing DNTs to perform real-time decision-making and maximize long-term rewards~\cite{zhou2023digital}. While DNTs serve as holistic simulators for predicting and optimizing network behavior, their effectiveness is limited without the integration of a distributed and continuous learning approach to address the scale, freshness, and privacy challenges of radio data. For instance, traditional centralized data processing often results in outdated twin models, bandwidth bottlenecks, and increased latency during inter-realm communications. By integrating FL and RL, DNTs synergistically combine secure, distributed learning with real-time optimization, enabling accurate predictions, adaptive policies, and scalable solutions for complex 6G networks. Existing articles have primarily focused on utilizing either DTs~\cite{9374645, almasan2022network} or ML techniques~\cite{kiran2021deep} to optimize network systems, but they fail to synergize these approaches into a collaborative framework for various downstream use cases. This article addresses that gap by pipelining these technologies in a highly adaptive and scalable system, capable of managing the complexities of 6G network environments with greater intelligence and efficiency.

Specifically, the synergy among DNTs, FL and RL introduces several technical challenges, particularly in maintaining reliability, real-time ability, and security across the communication system. 
One significant challenge is embedding controllable constraints into the RL framework while ensuring that the agent's autonomous decisions do not compromise physical network stability. Overly restrictive safety mechanisms can hinder exploration, reducing the system's ability to discover optimal solutions, while insufficient safeguards may expose the network to critical failures, such as service disruptions or security breaches. DNT plays a crucial role in addressing these challenges by simulating complex, real-world scenarios and monitoring novel conditions, enabling RL agents to undergo risk-informed training and receive timely interventions. However, achieving the balance between safety, exploration, and adaptability requires a carefully modular and well-structured design.
Besides, 6G networks require more than traditional ML models, as the complexity of these environments demands scenario-based forecasting rather than simple data-level predictions. FL can support both data forecasting and scenario mapping~\cite{safeRL}, but integrating FL with digital twinning for joint data-scenario forecasting presents challenges. A key difficulty is ensuring the DNT accurately mirrors the physical network while triggering ML models for predictions. Time-series forecasting depends on precise temporal and spatial data, and any discrepancies can lead to flawed predictions. In 6G scenarios, each distributed base station (BS) has unique attributes, such as user demand and hardware variations, which must be accurately captured in the DNT for precise network-wide forecasts.
Furthermore, 
when connecting DNT with multiple ML agents, the decentralized nature of the learning process introduces security risks, particularly from data and model poisoning attacks. In distributed 6G networks, adversaries can compromise the training process by injecting malicious data or altering model updates in both DNTs and physical networks, potentially leading to unsafe behaviors and decision-making.
This article addresses the above challenges associated with the synergy among DNTs, FL, and RL models, presenting for the first time comprehensive pipelines that demonstrate their practical applications in enhancing network optimization, performance forecasting, and system security. We also provide a case study using the proposed pipelines, offering experimental insights into several real-world utilities for 6G communication networks.

Our main contributions can be summarized as follows:
\begin{itemize}
    \item We propose a comprehensive DNT-empowered approach to enhance RL safety across diverse scenarios by integrating: (i) a safe RL mechanism leveraging DNT for reliable wireless resource optimization, and (ii) a defensive pipeline that protects federated RL-based safety-critical systems against model poisoning attacks. 
    \item We present an innovative distributed twinning framework for joint data-scenario forecasting, utilizing adaptive time-series mapping techniques for the efficient creation and continuous synchronization of DNTs.
    \item We validate our framework with two comprehensive case studies. The first evaluates the use of DNTs with safe RL for reliable edge caching, and the second examines the integration of secure federated RL with DNTs to improve the robustness of autonomous driving systems.
\end{itemize}

\section{Background and Preliminaries}

\subsection{Digital Network Twins}
Traditional digital twins often represent physical networks as multiple independent twins, making it challenging to achieve a holistic approach to network management. However, DNTs interconnect multiple task-specific twins, creating a unified framework that integrates data from diverse sources. This interconnected approach generates abundant, multidimensional data, enabling advanced analytics and optimization.
The development of a DNT involves several key stages, starting with data collection from the physical network, which includes parameters such as network topology, signal propagation characteristics, traffic flows, and device configurations. 
This data is then used to create a virtualized model of the physical realm, typically incorporating high-fidelity simulations that reflect both the dynamic and static aspects of network operations. 
Next, the virtual twin continuously synchronizes with the real-world network, ensuring that the digital model remains an accurate and up-to-date representation. 
This feedback loop between the physical and virtual networks allows for the simulation of various network conditions, the evaluation of potential optimizations, and even the detection of security vulnerabilities \cite{alcaraz2022digital}. 
DNTs are being promisingly adopted in 6G communication networks due to their ability to perform what-if analysis on system configuration and provide a safe testing ground for new protocols and networking strategies \cite{safeRL}. 
Recent studies have also examined the use of DNTs for enhancing network resilience, particularly in scenarios involving unexpected traffic surges, physical disruptions, or adversarial attacks \cite{zhang2024securing}.

\subsection{Federated Learning}
FL is a decentralized machine learning framework that allows multiple clients (such as edge devices or organizations) to collaboratively train a shared global model while keeping their local datasets private. 
The typical procedure of FL starts with the initialization of a global model at a central server, which is distributed to the participating clients. Each client independently trains the model on its local data for a predefined number of iterations. 
After local training, clients send only the updated model parameters or gradients, not the raw data, back to the central server, where these updates are aggregated (usually via weighted averaging) to update the global model. This process is repeated over multiple communication rounds until the global model converges \cite{mcmahan2017communication}. 
The primary challenge in FL lies in dealing with non-iid (non-independent and identically distributed) data across clients, which can lead to model convergence issues. 
Additionally, communication bottlenecks, privacy risks, and security vulnerabilities, such as poisoning attacks and model inversion attacks, have been widely discussed in the literature \cite{fang2020local}. 
To address these challenges, several methods have been proposed, including differential privacy, secure aggregation, and model compression techniques \cite{rodriguez2023survey}. FL has been applied across various 6G supported scenarios, including mobile edge computing, healthcare, and autonomous vehicles, providing a privacy-preserving alternative to traditional centralized learning.

\subsection{Reinforcement Learning}
RL is another learning paradigm in which an agent interacts with an environment in a sequential decision-making process. The agent observes the current state of the environment, selects an action based on the current policy, and receives feedback in the form of a reward signal. The goal of agents is to learn an optimal policy that maximizes the expected cumulative reward over time. RL process is formalized using Markov Decision Processes (MDPs), which define the environment in terms of states, actions, rewards, and transition probabilities. Q-learning is one of the fundamental RL algorithms, where the agent learns to estimate the expected utility of taking specific actions in given states, gradually improving its decision-making process \cite{prudencio2023survey}. More advanced techniques, such as policy gradient methods, directly optimize the policy itself rather than relying on value function estimation. With the advent of deep learning, RL algorithms are combined with deep neural networks to handle high-dimensional state spaces, leading to breakthroughs in areas such as autonomous driving, robotics, and game playing~\cite{kiran2021deep}. 
These advancements in data storage and processing have enabled DNTs to function as RL optimizers~\cite{cheng2024toward}. By leveraging these capabilities, DNTs offer a unified platform for precise and reliable decision-making, supported by abundant generated data for RL agent training and predictive analytics.

\section{Challenges and Opportunities}

\begin{figure*}[htbp]
    \centering
    \includegraphics[width=0.98\textwidth]{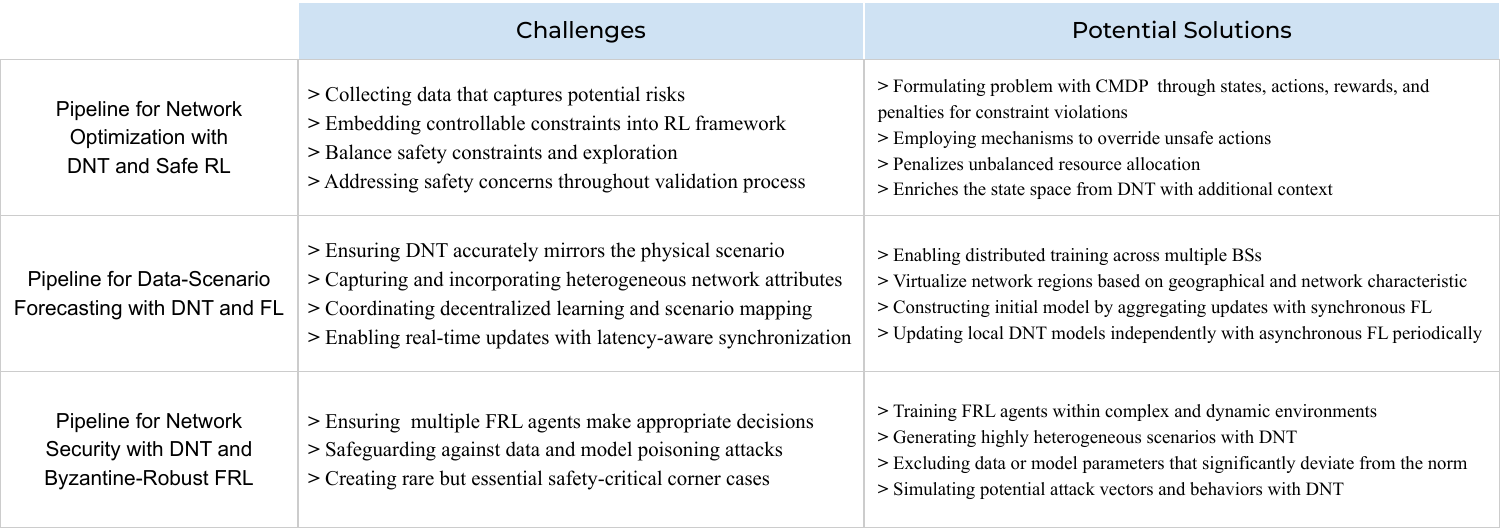}
    \vspace{-0.1in}
    \caption{Overview of challenges and potential solutions in the synergy of DNT and AI.}
    \label{fig:challenges}
\end{figure*}

In this section, we introduce the challenges and opportunities behind the synergy of DNT and artificial intelligence, as summarized in Fig.~\ref{fig:challenges}.

\subsection{Pipeline for Network Optimization with DNT and Safe RL}

In 6G networks, the growing demand for reliable, high-performance infrastructure presents challenges that traditional optimization methods are increasingly unable to address, as these networks will not only provide connectivity but also support critical systems such as healthcare, autonomous vehicles, and industrial automation, where even minor disruptions in resource allocation could lead to severe consequences. Addressing these concerns requires advanced solutions for intelligent decision-making within a framework that ensures both system stability and adaptability. RL, particularly when integrated with safety modules, offers a promising solution by enabling safe resource optimization in real-time, while adhering to system constraints. However, for RL to be effectively deployed in high-stake scenarios, it must be trained using comprehensive and diverse data that captures potential risks. This is where DNT becomes essential, providing the capability to simulate complex, real-world ``what-if'' scenarios, generating a wide range of case studies, and continuously monitoring critical conditions. By doing so, DNT allows RL agents to undergo rigorous, risk-informed training, ensuring their performance is robust, scalable, and adaptable to the dynamic nature of future changes.

Nevertheless, the integration of DNT and RL with safety modules introduces several complex challenges. One of the most significant is the difficulty of embedding controllable constraints into RL framework within a up-to-date virtual environment, ensuring that the agent can make autonomous decisions without compromising physical network stability. In highly dynamic environments, RL agents are inclined to explore new strategies, which can be risky without appropriate safeguards. If safety modules impose overly strict limitations, they risk stifling the agent’s capacity for exploration, potentially reducing its ability to discover optimal policies. Conversely, insufficient safety mechanisms may expose the network to critical failures, especially in time-sensitive or high-risk scenarios. Another challenge is ensuring consistent system performance, both during safe, constrained decision-making and during periods of exploration. Achieving a balance between safety and exploration is crucial for maintaining operational efficiency while preventing harmful actions. Lastly, addressing safety concerns throughout the validation process is paramount, as agents often encounter novel, untested states that pose significant risks. DNT plays a vital role in managing these risks by simulating and monitoring a large number of unseen conditions, enabling timely interventions to prevent unsafe decisions. This continuous oversight ensures that the system remains robust and generalizable, ultimately facilitating its deployment in real-world 6G networks where system safety and adaptability are of the utmost importance.

\begin{figure}[htbp]
    \centering
    \includegraphics[width=0.33\textwidth]{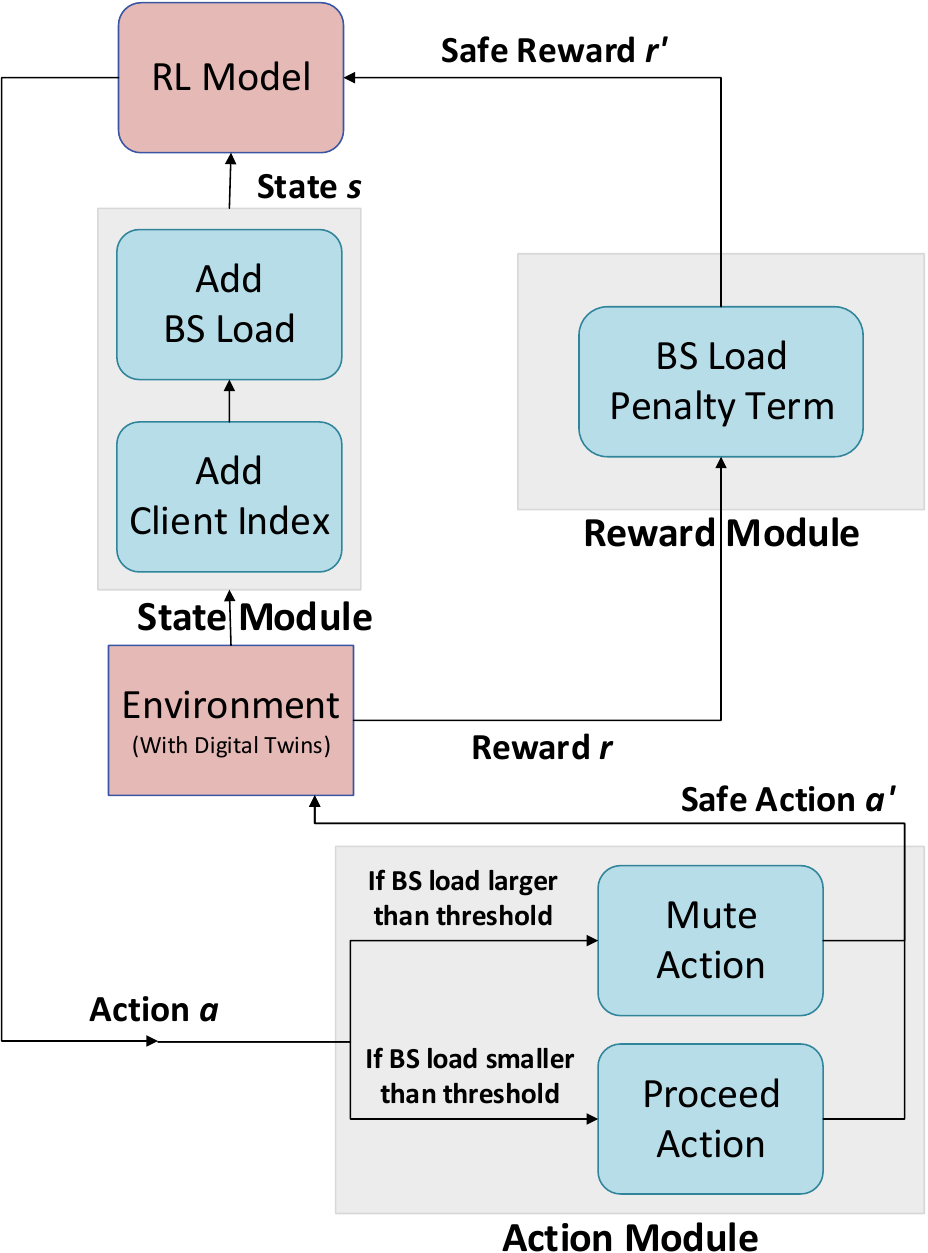}
    \vspace{-0.1in}
    \caption{A close-up view of the pipeline. The gray box highlights the active area of the safety-oriented intervention modules.}
    \label{fig:module}
\end{figure}

As shown in Fig.~\ref{fig:module}, one potential pipeline for deploying DNT-enabled RL for resource allocation involves integrating a Constrained Markov Decision Process (CMDP) optimizer, along with multiple safety-enhancing learning modules. At its core, the \textit{CMDP module} formulates the resource management problem by capturing the dynamic behavior of the network through states, actions, rewards, and penalties for constraint violations. This structure ensures that resource constraints are embedded into an RL framework.
Note that, the RL systems integrate both exploration and exploitation to learn optimal policies. Exploration enables an agent to sample new actions and discover potentially beneficial behaviors, while exploitation uses the current policy to maximize rewards based on known information. However, excessive or unchecked exploration can result in erratic, unreliable actions. 
To mitigate these risks, it is essential to implement robust monitoring and safety measures during both the exploration and exploitation phases. For instance, incorporating constraint enforcement, risk-aware decision-making, and fail-safe protocols can help ensure that the agent operates within safe boundaries. 
One strategy, such as the epsilon-greedy policy, gradually decreases the exploration rate as the agent gains confidence in its learned policy. While this approach provides a straightforward mechanism for balancing exploration and exploitation, additional safety interventions are crucial to ensure reliable decision-making.
Inside the RL framework, the \textit{DNT module} serves as a real-time virtual replica of the physical network, continuously synchronized with it, and generates future network states—such as user demands and BS loads—which are fed into the CMDP module. By providing predictive insights, DNT enhances the RL agent's ability to adapt to fluctuating conditions and helps ensure consistent performance in dynamic environments.
The RL agent, based on a Deep Q-Network (DQN) as an example, leverages the predictions from the CMDP and DNT modules to learn optimal resource allocation policies.

To further safeguard the system, the pipeline can incorporate multiple intervention modules. For instance, the \textit{State Intervention Module} enriches the state space from the DNT with additional context, such as risky BS loads, to improve decision-making. The \textit{Action Intervention Module} implements backup policies that override unsafe actions, preventing potential network overloads or performance degradation. Lastly, the \textit{Reward Intervention Module} penalizes unbalanced resource allocation, encouraging better load distribution and ensuring system stability. These intervention modules create a safety-driven environment that balances exploration and performance optimization. On another note, the DNT can also function as a real-time monitor, identifying and correcting any risky actions promptly, thereby making the pipeline self-adaptive and closed-loop. Particularly, the intervention modules can be integrated into other state-of-the-art RL algorithms, such as Deep Deterministic Policy Gradient (DDPG) and Proximal Policy Optimization (PPO), potentially enhancing network management performance beyond that achieved by DQN.
A case study on edge caching problem to evaluate the performance of this proposed pipeline is demonstrated in Sec. IV-B.

\subsection{Pipeline for Data-Scenario Forecasting with DNT and FL}

6G networks will require more than just traditional ML models because the complexity and dynamism of environments demand scenario-based forecasting rather than simple data-level prediction. With massive data traffic, multi-modal services, and real-time, highly variable conditions (e.g. mobility, user density, and network configurations), ML alone cannot adapt quickly enough. By integrating a DNT with ML, 6G networks can foresee various future scenarios—such as network-wide congestion or infrastructure changes—and optimize system configuration accordingly. 
However, combining DNT with ML for time-series data scenario forecasting presents several technical challenges.
These include ensuring real-time synchronization of both environmental and radio data between the physical network and the DNT, accurately modeling complex scenarios such as user behavior and network load, as well as managing the high computational demands for large-scale simulations. Additionally, integrating multi-modal data streams while adapting to the distributed nature of 6G deployments, further complicate the implementation of this technical integration.

As a remedy, FL provides a powerful solution by enabling distributed training across multiple BSs, each contributing localized data to build a more comprehensive global model, e.g., for scenario generation. 
Consider the FL-based DNT applied to wireless traffic prediction problems. Each BS locally generates traffic predictions by analyzing historical data and contributing to a global twin model without sharing raw data, ensuring privacy. The system iteratively optimizes the global model to minimize prediction errors across all BSs, improving its ability to forecast traffic and understand network behavior.

Despite this promise, pipelining an FL process and digital twinning for joint data-scenario forecasting introduces several technical challenges. One of the foremost difficulties is ensuring that the DNT accurately mirrors the physical scenario while initiating the ML module for data-level predictions. Since time-series forecasting heavily relies on accurate temporal and spatial information, any discrepancies in how the physical environment is represented in the twin can lead to inaccurate data predictions.
Additionally, each BS in the network possesses unique attributes, including user demand patterns, hardware variations, and environmental conditions. Capturing and incorporating these heterogeneous factors into the DNT is also critical for generating precise forecasts across the network.
Besides, jointly coordinating decentralized learning and scenario mapping across regional BSs, each with its own local data and computational resources, is a non-trivial task, particularly when the data attributes of each local area vary significantly. Ensuring that all BSs contribute effectively to the global model without creating bottlenecks or inefficiencies is crucial. Achieving this requires careful management of data transmission, model aggregation, and local updates to maintain the overall system's performance without compromise.
Furthermore, the multi-scale temporal nature of network operations adds another layer of complexity. For instance, certain tasks require real-time updates, while others can tolerate delays of several milliseconds or more. Latency-aware updates are essential to accurately reflect current network conditions and avoid outdated data leading to suboptimal decisions at the appropriate frequency. Maintaining multiple levels of synchronization requires adaptive data processing and updates of the DNT, which is a technically challenging task given the need to handle large volumes of data without compromising the accuracy of time-series forecasting. 

\begin{figure}[htbp]
    \centering
    \includegraphics[width=0.48\textwidth]{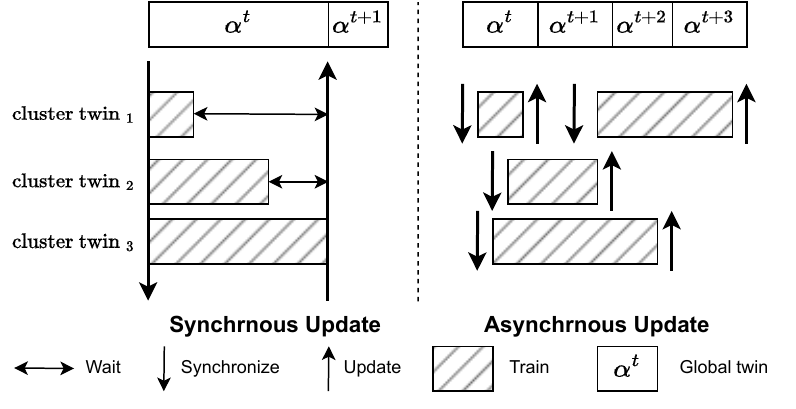}
    \vspace{-0.1in}
    \caption{Comparison between synchronous and asynchronous FL.}
    \label{fig:AFL}
\end{figure}

To this end, one potential solution is to leverage a combination of synchronous and asynchronous FL for jointly scenario twinning and data forecasting, as depicted in Fig.~\ref{fig:AFL}.
In the first stage, a periodic clustering mechanism virtualizes network regions based on geographical and network characteristics, creating local DNTs and a global DNT model. The algorithm dynamically evaluates factors like geographical proximity, backhaul capacity, coverage overlap, and traffic patterns to form clusters. Two possible methods are used, that one ensures a fixed number of clusters by iteratively removing weak connections, similar to Girvan-Newman, while the other adapts using a Louvain-like approach to maximize modularity. These clusters enable efficient network region virtualization and global twin updates.

In the second stage, \textit{synchronous} FL plays a critical role by aggregating updates from all BSs (clustered into several local DNTs) in a coordinated manner, ensuring that the global DNT model is built with high consistency and accuracy. This approach is crucial during the early stages of model construction and data forecasting, where uniformity is needed to create a robust representation of the network. It is essential during this phase to capture the unique attributes of each BS, such as user demand patterns, hardware variations, and environmental conditions, ensuring that these heterogeneous factors are integrated into the global DNT. This careful aggregation mitigates the risk of misalignment between the physical network and its DNT, thereby enhancing the precision of following time-series forecasts.
As the system progresses and the need for real-time responsiveness becomes necessary, the framework transitions to an \textit{asynchronous} FL approach. This shift allows each local DNT to update its local model independently, accommodating the varying conditions and dynamic nature of each BS cluster without requiring global synchronization. This flexibility significantly enhances the system's ability to adapt to network changes in real-time. In essence, such asynchronous process reduces bottlenecks that could arise from synchronous updates, allowing faster model updates and minimizing delays. 
In essence, synchronous FL ensures consistency by requiring all participating cluster twins to submit updates in each communication round, but it is sensitive to latency from stragglers with slower communication or limited computation. This can be mitigated by allowing partial cluster twin participation, where only a subset of updates is aggregated per round. Conversely, asynchronous FL reduces communication overhead and latency bottlenecks by letting cluster twins update the global twin at different times. However, it risks stale contributions from outdated models, which can be managed using staleness-aware weighted aggregation to maintain global model accuracy.
To further maintain the model freshness, each local DNT is periodically updated with real-time data from its clustered BSs, and clusters are \textit{reformed} based on key network attributes, such as backhaul link capacity and traffic distribution. 
By employing such a combined clustering mechanism with synchronous and asynchronous FL-based pipeline, the DNT remains accurate and responsive, enabling effective real-time scenario monitoring, data forecasting, and decision-making in complex 6G network environments.

\subsection{Pipeline for Network Security with DNT and Byzantine-Robust FRL}

In 6G networks, the integration of federated RL (FRL) with DNT opens new possibilities for improving the efficiency and security of decision-making systems. 
Single-agent RL struggles with low sampling efficiency in safety-critical domains like autonomous driving, where addressing rare edge cases is vital to ensure safety. FRL mitigates these challenges by enabling decentralized agents to collaborate, improving learning efficiency, generalization, and data privacy in dynamic environments. While both FL and FRL use distributed training, FL relies on static local datasets for pattern recognition, whereas FRL involves agents learning through interaction with environments. FL assumes IID data and convex loss functions for theoretical guarantees, but FRL faces added complexity due to trajectory heterogeneity and non-convex losses, complicating convergence analysis.
At its core, the DNT plays a crucial role by generating highly heterogeneous or attack scenarios that reflect the complexity of the engaged environments, ensuring that FRL agents are trained on a diverse range of vulnerable cases.

However, integrating DNT with FRL introduces a range of challenges. One of the foremost difficulties is ensuring that multiple RL agents make appropriate decisions in scenarios simulated by DNT, 
where incorrect actions from one or more agents could lead to severe consequences or chain failures. Although DNT provides a controlled platform to simulate these multi-view situations, the challenge lies in ensuring that all agents can transfer their learning to shared environments and handle unforeseen conditions effectively. 
Moreover, the decentralized nature of FRL poses security risks, particularly in safeguarding against data and model poisoning attacks. In a distributed 6G networks, adversaries can attempt to corrupt the training process by injecting malicious data or altering model updates, potentially leading to unsafe behaviors. To address these risks, the FRL framework must incorporate robust security measures, such as secure aggregation techniques and anomaly detection systems, to protect against adversarial threats such as Byzantine attacks. DNT is instrumental here, as it can generate various attack vectors, allowing operators to identify and mitigate vulnerabilities before they impact real-world operations.
Particularly, DNT’s ability to create rare but essential corner cases further strengthens the training process. These scenarios, such as high-speed collision avoidance or complex multi-agent interactions in autonomous driving, are critical for developing the system's ability to handle extreme and infrequent events. However, designing these corner cases presents its own set of challenges, as they must strike a balance between realism and complexity. The scenarios from DNT that are too unrealistic could distort the learning process, while overly simplistic ones may not sufficiently challenge the system. Achieving this balance is crucial for ensuring that the FRL model is both robust and capable of performing effectively in real-world security applications.

To synergize FRL and DNT as a pipeline for security-sensitive environments, one approach is to integrate DNT with a secure model aggregation mechanism within the FRL framework. This allows multiple agents to collaborate, enhancing both sampling efficiency and security. The DNT module first generates highly heterogeneous scenarios, providing a diverse set of training data that mirrors the complexity and dynamism of real-world environments. Subsequently, a server-based filtering mechanism can be implemented to exclude data or model parameters from agents that significantly deviate from the norm, ensuring that only consistent and benign inputs are used for model aggregation.
This secure aggregation technique functions as an anomaly detection engine, ensuring that only trustworthy updates from agents contribute to the global model. In parallel, the DNT module can simulate potential attack vectors, such as data poisoning or adversarial examples, enabling early detection and mitigation of vulnerabilities. We will demonstrate the effectiveness of this pipeline through a case study in Sec. IV-C, showing how it enhances both security and performance in complex 6G network environments.

\section{Case Studies}

In this section, we provide two case studies derived from the proposed pipelines discussed in Sec. III. The first one assesses the effectiveness of integrating DNTs with a safe RL framework in edge caching scenarios, while the second examines the performance of our DNT-FRL pipeline in securing autonomous vehicular systems.

\subsection{DNT-enabled RL for Safe Caching in 6G Networks}

In this part, we conduct a case study based on the proposed pipeline in Sec.~III-A. The experimental setup is to optimize cache replacement in a wireless network. The evaluations are conducted on five BSs, each with a cache of 150 slots, serving multiple clients. In this setup, BSs independently manage their caches without sharing data with neighboring stations, making the optimization problem more complex. The state space includes information such as BS index, time since a content item was last cached, and frequency of requests for cached items. Additionally, the state space is expanded to consider the current load on each BS and the client making the request, allowing RL model to make informed caching decisions. The action space involves deciding whether to accept a request and determining which cache slot to use. The reward function incentivizes cache hits and penalizes cache misses and BS overloads.

We integrate the constructed DNT into RL optimization framework, where it functions as both a RL optimizer and a safeguard. Real-time synchronization is achieved through two-way communication between the physical network and the DNTs, facilitated by the proposed methods in Sec. III-B. The twinning process in Sec. III-B supplies the DNTs with historical caching data, including details such as requests, content types, frequencies, and other relevant attributes. Using data-driven techniques like long short-term memory (LSTM), the DTs generate one-step forecasts for upcoming content requests. These forecasts allow for the analysis of content occurrence distributions. Historical caching datasets spanning both common and rare wireless scenarios can be utilized to train DNTs, producing content distributions based on their likelihood of occurrence. The generated content and associated attributes serve as inputs to the RL model where they act as the state representations within the RL algorithm. Following the RL model's decision-making process, real-time data is fed back to the DNTs for further execution, ensuring the twinning model remains accurate in a closed-loop system.

The dataset used in the evaluation is generated from real-world patterns, modeled using the Zipf distribution with \(p = 0.8\), which represent the skewed nature of content popularity. 
Several intervention modules are integrated into the RL framework to ensure network safety. The state intervention module interfaced with DNT environments extends the state space to include the BS load, allowing the model to proactively manage traffic distribution. The action intervention module overrides RL decisions that could lead to BS overload by muting unsafe actions when necessary. Furthermore, the reward intervention module modifies the reward function to prioritize load balancing across the BSs by introducing a penalty for imbalances. 

Performance is evaluated using key metrics, including cache hit rate, action intervention rate, and BS load balance. The cache hit rate measures the percentage of requests successfully served from the cache, indicating the efficiency of the caching strategy. 
The intervention rate measures the proportion of actions intercepted to prevent invalid or unsafe replacements, serving as an indicator of the effectiveness of the action intervention module in the proposed approach. Lower intervention rates suggest greater safety of the RL actions.
BS load balance is crucial in assessing how evenly traffic is distributed across the network, with the goal of avoiding BS overload. 

\begin{figure}[htbp]
    \centering
    \includegraphics[width=0.48\textwidth]{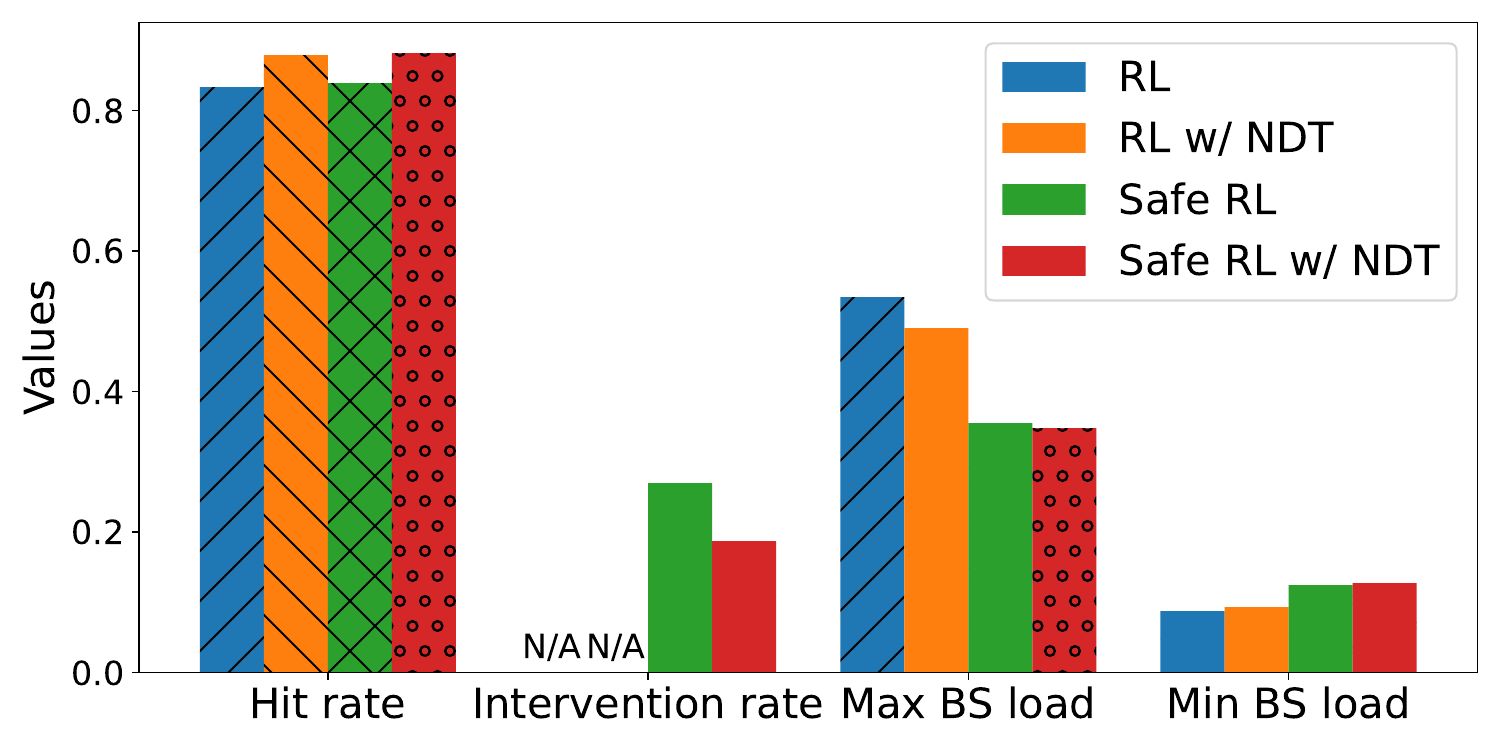}
    \vspace{-0.1in}
    \caption{Safe edge caching with synergy of RL and DNT.}
    \label{fig:cache}
\end{figure}

Fig.~\ref{fig:cache} illustrates the performance of different RL and DNT configurations across key metrics. 
The baseline RL model, without the DNT integration, shows moderate performance with a cache hit rate of around 0.82, while the maximum and minimum BS loads are around 0.53 and 0.08, indicating some imbalance in load distribution among BSs. When the DNT module is integrated in the pipeline, the system's load balancing improves, as reflected by a slight decrease in both maximum and minimum BS loads. This suggests that DNT helps RL model better manage the load distribution across BSs by utilizing real-time data for more accurate predictions. With the addition of safety intervention modules, the system achieves a significant increase in cache hit rate, reaching over 0.84, while maintaining a higher intervention rate near 0.35, which indicates that the system actively intervenes to ensure cache safety. The maximum BS load decreases, reflecting more even traffic distribution across BSs, which helps prevent overloads. 
The optimal configuration, combining intervention modules with DNT and RL, achieves the highest hit rate, close to 0.88, while further reducing the maximum BS load to around 0.35 and improving the minimum BS load to 0.12. These results demonstrate the effectiveness of our proposed pipeline in enhancing both cache efficiency and load balancing in the network system.

\subsection{DNT-enabled FRL for Securing Internet of Vehicles}

This case study evaluates the performance of our proposed pipeline within a DNT designed for FRL in autonomous driving scenarios, based on the proposed pipeline in Sec.~III-C. Our developed DNT environment, referred to as \textit{HighwayDT}~\cite{NSFOAC_GitHub}, simulates highway driving conditions in real-time, modeling vehicle dynamics and interactions to train and validate multiple RL agents under complex and high-risk scenarios. The agents are responsible for controlling ego vehicles in three-car following scenarios, where they must balance longitudinal control and collision avoidance. 
Each agent is trained locally within its DNT instance, and the collective knowledge is aggregated using an FRL framework to improve decision-making in various driving scenarios. The reward function penalizes collisions and rewards successful maneuvers, ensuring the agents optimize for safety and performance.

The DNT supports parallel scenario generation, enabling RL agents to learn in a range of dynamic and heterogeneous driving conditions, including different vehicle behaviors.
The evaluation is conducted across over 50,000 generated scenarios, where the primary metric is the no-collision rate, which measures the percentage of scenarios in which no collisions occur. This metric reflects the effectiveness of RL agents in making safe driving decisions under various conditions, including those designed to simulate extreme or unexpected events. FRL environment is also tested against multiple poisoning attacks during the data transmission, where a portion of the agents is controlled by adversaries. These malicious agents attempt to disrupt the decision-making process, while the FRL system is evaluated by comparing no-collision rates under different attack strategies.

\begin{figure}[htbp]
    \centering
    \includegraphics[width=0.48\textwidth]{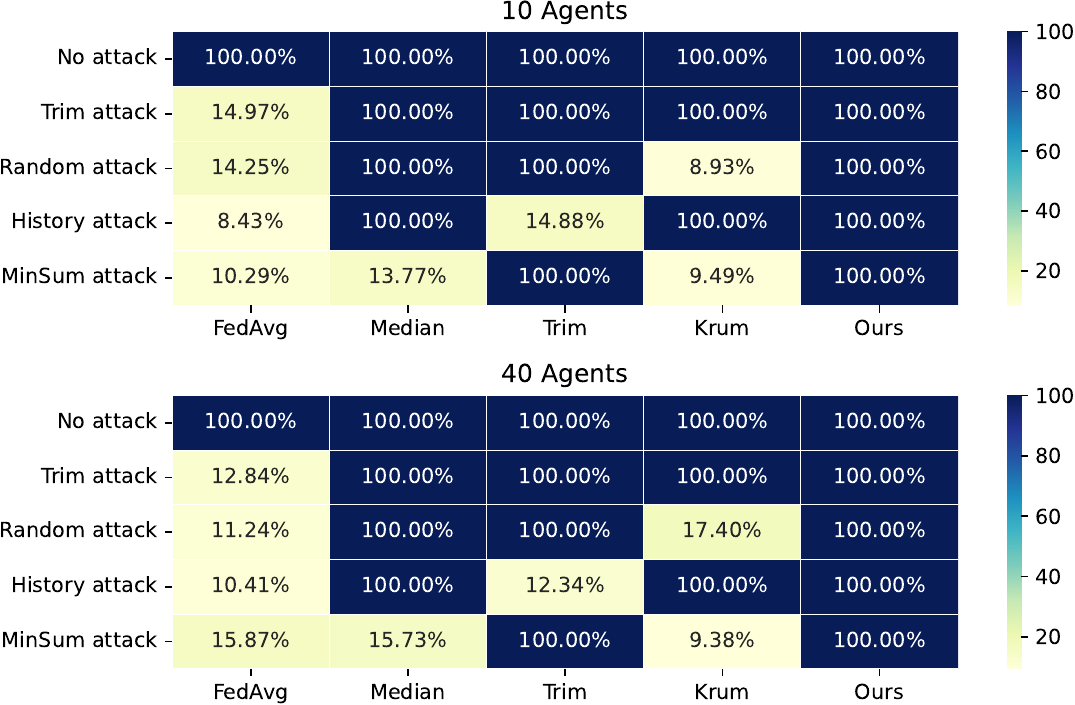}
    \vspace{-0.1in}
    \caption{Secure Internet of Vehicles with FRL trained and tested within DNTs.}
    \label{fig:iov}
\end{figure}

Fig.~\ref{fig:iov} presents a heatmap comparing the no-collision rates achieved by various secure aggregation rules across multiple attack scenarios, highlighting the robustness of each method. In the absence of attacks, all aggregation rules achieve a perfect no-collision rate of 100\%, indicating that RL agents can safely control the vehicles without interference. However, under adversarial conditions, performance varies significantly across the different baseline rules. Under various attacks, our method maintains robustness with a 100\% no-collision rate, thanks to the embedded secure aggregation and DNT modules, which provide data-level anomaly detection and informed adversarial experience.
It also demonstrates consistent performance across varying numbers of agents, achieving a 100.0\% no-collision rate in both the 10-agent and 40-agent scenarios, highlighting the potential scalability of our proposed pipeline for deployment in large-scale systems.
The overall performance across various attack types demonstrates our pipeline's ability to ensure the safety and stability of RL agents in adversarial settings.

\section{Conclusion}

Synergizing DNT with FL and RL offers a novel approach for managing and optimizing complex 6G networks. DNTs provide real-time simulations and monitoring, while FL ensures decentralized, privacy-preserving model training, and RL enables adaptive decision-making in dynamic environments. This synergy enhances network resilience, efficiency, and scalability. 
This article proposes several pipelines for integrating DNT with ML techniques within a coherent framework, offering a secure and adaptive solution for network optimization, forecasting, and security. 
Through experimental validation, these approaches demonstrate real-world applicability, positioning DNTs as key tools for future network components.


\bibliographystyle{IEEEtran}
\bibliography{refs}

\vfill

\vspace{11pt}

\end{document}